\documentclass[twocolumn,10pt]{IEEEtran}
\usepackage{ifpdf, flushend}

\ifCLASSINFOpdf
  \usepackage[pdftex]{graphicx}
  \graphicspath{{../pdf/}{../jpeg/}}
  \DeclareGraphicsExtensions{.pdf,.jpeg,.png}
\else
  \usepackage[dvips]{graphicx}
  \graphicspath{{../eps/}}
  \DeclareGraphicsExtensions{.eps}
\fi
\usepackage[cmex10]{amsmath}
\usepackage {amssymb}
\usepackage{algorithmic}
\usepackage{array}
\usepackage{mdwmath}
\usepackage{mdwtab}
\usepackage{eqparbox}
\usepackage{algorithm}
\usepackage{algorithmic}

\newcommand{\argmax}{\operatornamewithlimits{argmax}}
\newcommand{\beq}{\begin{equation}}
\newcommand{\eeq}{\end{equation}}
\newcommand{\beqn}{\begin{eqnarray}}
\newcommand{\eeqn}{\end{eqnarray}}
\newcommand{\beqno}{\begin{eqnarray*}}
\newcommand{\eeqno}{\end{eqnarray*}}
\newcommand{\bma}{\begin{displaymath}}
\newcommand{\ema}{\end{displaymath}}
\newcommand{\bnu}{\begin{enumerate}}
\newcommand{\enu}{\end{enumerate}}
\newcommand{\bce}{\begin{center}}
\newcommand{\ece}{\end{center}}
\newcommand{\btb}{\begin{tabular}}
\newcommand{\etb}{\end{tabular}}

\begin{document}
%
\title{General Analytical Framework for Cooperative Sensing and Access Trade-off Optimization}

\author{\IEEEauthorblockN{Le Thanh Tan and Long Bao Le}  
\thanks{The authors are with INRS-EMT, University of Quebec,  Montr\'{e}al, Qu\'{e}bec, Canada. 
Emails: \{lethanh,long.le\}@emt.inrs.ca. }}

\maketitle

\begin{abstract}
\boldmath
In this paper, we investigate the joint cooperative spectrum sensing and access design problem for multi-channel cognitive radio networks.
A general heterogeneous setting is considered where the probabilities that different channels are available, SNRs of the signals
received at secondary users (SUs) due to transmissions from primary users (PUs) for different users and channels can be different.
We assume a cooperative sensing strategy with a general a-out-of-b aggregation rule and design a synchronized MAC protocol
so that SUs can exploit available channels. We analyze the sensing performance
and the throughput achieved by the joint sensing and access design. Based on this analysis, we develop algorithms
to find optimal parameters for the sensing and access protocols and to determine channel assignment for SUs to
maximize the system throughput.
Finally, numerical results are presented to verify the effectiveness of our design and demonstrate the relative performance of
our proposed algorithms and the optimal ones.
\end{abstract}

\IEEEpeerreviewmaketitle

\section{Introduction}

Design and analysis of MAC protocols for cognitive radio networks is an important research topic.
There has been growing literature on this topic over the last few years  \cite{Cor09} and \cite{Konda08} (see \cite{Cor09} 
for a survey of recent works). 
However, most existing works either assume perfect spectrum sensing or do not explicitly model the sensing imperfection in their
 design and analysis. In \cite{Le11}, we design and optimize the sensing and MAC protocol parameters where each SU is assumed to perform parallel
 sensing on all channels and it can use all available channels for data transmission. This can be considered as an extension 
 of throughput-sensing optimization framework of \cite{R1} from the single-user to the multi-user setting. 
In \cite{Le12}, we consider a scenario where each SU can exploit at most one channel for transmission.
All these works do not consider cooperative sensing and its design issues.

Cooperative spectrum sensing has been shown to  improve the sensing performance \cite{Gan07}--\cite{Seu10}. In a cooperative sensing strategy, 
each SU performs sensing independently and then sends its sensing results to an access point (AP). The AP then makes decisions on the idle/busy
status of each channel by using certain aggregation rule. In \cite{Quan08}, weighted data based fusion is proposed to improve 
sensing performance. In \cite{Peh09}-\cite{Wei09}, the optimization of cooperative sensing using an a-out-of-b rule is performed.
 In \cite{Wei11}, the game-theoretic based method is taken to develop a cooperative spectrum sensing strategy.
However, these works only focus on design and optimization of cooperative sensing without considering the spectrum access
problem (i.e., how SUs share the available spectrum). Furthermore, these sensing optimization works are performed for a single channel and
 homogeneous scenario where channel parameters such as SNRs, probabilities that different channels are available are the same. 
In \cite{Seu10}, the authors investigate a multi-channel scenario where each SU simultaneously senses all channels using one receiver per channel
 and calculates the log-likelihood ratio of observed measurement. Then AP collects these statistics to decide when to terminate the process. 
All of these existing works do not consider the joint cooperative sensing and access design under the heterogeneous
setting.

In this paper, we propose the general cooperative sensing-access framework for the non-homogeneous scenario
where a general a-out-of-b aggregation rule is assumed at the AP.
Specifically, the contributions of this paper can be summarized as follows: 
\emph{i}) we design joint cooperative sensing and synchronized MAC protocols for a multi-channel cognitive radio network.
We derive the spectrum sensing performance for a-out-of-b aggregation rule and 
we perform the throughput analysis of our proposed sensing and access design.
\emph{ii}) we propose solutions for two parameter optimization problems of our proposed design.
Specifically, given a channel assignment, we study how to determine the sensing time and 
contention window of the MAC protocol. Moreover, we consider the channel assignment problem
for throughput maximization where we present both brute-force search optimal algorithm
and the low-complexity greedy algorithm.
\emph{iii}) we present numerical results to illustrate the performance of the proposed MAC protocols and the throughput gains due 
to optimized design compared to the non-optimized one.

The remaining of this paper is organized as follows. Section~\ref{SystemModel} describes the system model, sensing, and access design. 
Throughput analysis, optimization of spectrum sensing, access, and channel assignment are performed in Section~\ref{MultipleChan}.
Section~\ref{Results} presents numerical results followed by concluding remarks in Section~\ref{conclusion}.
 
\section{Spectrum Sensing and Access Design}
\label{SystemModel}

In this section, we describe the system model, spectrum sensing, and access design for the cognitive radio networks.

\subsection{System Model}
\label{System}

\begin{figure}[!t] 
\centering
\includegraphics[width=60mm]{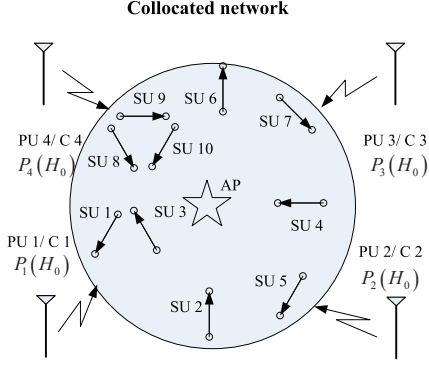}
\caption{Network model (PU: primary user, SU: secondary user)}
\label{Fig1}
\end{figure}

We consider a network setting where $N$ pairs of secondary users (SUs) opportunistically exploit available
frequency bands in $M$ channels for data transmission. For simplicity, we refer to 
pair $i$ of SUs simply as SU $i$. 
We assume that each SU can exploit multiple available channels for transmission (e.g., by using
OFDM technology). We will design a synchronized MAC protocol for channel access. We assume 
that each channel is either in the idle or busy state for each
predetermined periodic interval, which is called a cycle in this paper.

We further assume that each pair of SUs can overhear transmissions from other pairs of SUs (i.e., collocated networks). 
There are $M$ primary users (PUs) each of which may or may not use one corresponding channel for its data transmission
in any cycle. In addition, it is assumed that transmission from any pair of SUs on a particular channel will affect the primary receiver
which receives data on that channel. The network setting under investigation is shown in Fig.~\ref{Fig1}.

\subsection{Cooperative Spectrum Sensing}
\label{Ss} 

We assume that each SU $i$ is assigned in advance a set of channels $S_i$ where
it senses all channels in this assigned list at beginning of each cycle.
Optimization of such channel assignment will be considered in the next section.
Upon completing the channel sensing, each SU $i$ sends the idle/busy states
of all channels in $S_i$ to the access point (AP) for further processing. 
The AP upon collecting sensing results from all SUs will decide idle/busy
status for all channels. Then, the AP broadcasts the list of available channels to all SUs. 
SUs are assumed to rely on a distributed MAC protocol to perform access resolution where
the winning SU transmits data by using all available channels.
Detailed MAC protocol design will be elaborated later.

Let $\mathcal{H}_0$ and $\mathcal{H}_1$ denote the events that a particular PU is idle and active, respectively
(i.e., the corresponding channel is available and busy, respectively) in a cycle. In addition, let 
 $\mathcal{P}_j \left( \mathcal{H}_0 \right)$ and  $\mathcal{P}_j \left( \mathcal{H}_1 \right) = 1 - \mathcal{P}_j \left( \mathcal{H}_0 \right)$ be the probabilities that channel $j$ is available and not available for all SUs, respectively.
We assume that SUs employ an energy detection scheme and let $f_s$ be the sampling frequency used in the
sensing period for all SUs. There are two important performance measures,
which are used to quantify the sensing performance, namely detection and false alarm probabilities. In particular,
detection event occurs when a SU successfully senses a busy channel and false alarm
represents the situation when a spectrum sensor returns a busy status for an idle channel (i.e., a transmission opportunity
is overlooked).

Assume that transmission signals from PUs are complex-valued PSK signals while the noise at the SUs is independent and identically distributed circularly symmetric complex Gaussian $\mathcal{CN}\left( {0,{N_0}} \right)$ \cite{R1}. Then, the
detection and false alarm probabilities for the channel $j$ at SU $i$ can be calculated as \cite{R1}
\beqn
\label{eq1}
\mathcal{P}_d^{ij}\left( \varepsilon ^{ij} ,\tau^{ij}  \right) = \mathcal{Q}\left( \left( \frac{\varepsilon ^{ij} }{N_0} - \gamma ^{ij}  - 1 \right)\sqrt {\frac{\tau^{ij} f_s}{2\gamma ^{ij}  + 1}}  \right), 
\eeqn
\beqn
 \mathcal{P}_f^{ij}\left( \varepsilon ^{ij} ,\tau^{ij}  \right) = \mathcal{Q}\left( \left( \frac{\varepsilon ^{ij} }{N_0} - 1 \right)\sqrt {\tau^{ij} f_s}  \right) \hspace{2.5cm} \nonumber \\ 
 = \mathcal{Q}\left( \sqrt {2\gamma ^{ij}  + 1} \mathcal{Q}^{ - 1}\left( \mathcal{P}_d^{ij}\left(  \varepsilon ^{ij} ,\tau^{ij}   \right) \right)+\sqrt {\tau^{ij} f_s} \gamma ^{ij}  \right),  \label{eq2}
\eeqn
where $i \in \left[ {1,N} \right]$ is the index of a SU link, $j \in \left[ {1,M} \right]$ is the index of a channel, ${\varepsilon ^{ij}} $ is the detection threshold for an energy detector, ${\gamma ^{ij}} $ is the signal-to-noise ratio (SNR) of the PU's signal at the SU, $f_s$ is the sampling frequency, $N_0$ is the noise power, $\tau^{ij}$ is the sensing interval of SU $i$ on channel $j$, and $\mathcal{Q}\left( . \right)$ is defined as $\mathcal{Q}\left( x \right) = \left( {1/\sqrt {2\pi } } \right)\int_x^\infty  {\exp \left( { - {t^2}/2} \right)dt}$. 

We assume that a general cooperative sensing scheme, namely $a$-out-of-$b$ rule, is employed by the AP to determine the idle/busy status
of each channel based on reported sensing results from all SUs. Under this scheme, the AP will declare that a channel is busy
if $a$ or more SUs out of $b$ SUs report that the underlying channel is busy. 
The a-out-of-b rule covers different other rules including OR, AND and Majority rules as special cases. In particular, when $a=1$, it is OR rule; when $a=b$, it is AND rule; and when $a=\left\lceil b/2\right\rceil$, it is Majority rule. 
Let consider channel $j$. Let $\mathcal{S}_j^U$ denote the set of SUs that sense channel $j$ and $b_j=\left|\mathcal{S}_j^U\right|$
be the number of SUs sensing channel $j$.
Then the AP's decision on the status of channel $j$ will result in detection and false alarm probabilities for this channel, which can be
 calculated as, respectively \cite{Wei11}
\beqn
\mathcal{P}_d^j\left( {\vec \varepsilon ^j}, {\vec \tau^j} , a_j  \right) = \sum_{l=a_j}^{b_j} \sum_{k=1}^{C_{b_j}^l} \prod_{i_1 \in \Phi^k_l} \mathcal{P}_d^{i_1j} \prod_{i_2 \in \mathcal{S}_j^{U} \backslash \Phi^k_l} \mathcal{\bar P}_d^{i_2j} \label{eq1_css_1}\\
\mathcal{P}_f^j\left( {\vec \varepsilon ^j} , {\vec \tau^j}, a_j  \right) = \sum_{l=a_j}^{b_j} \sum_{k=1}^{C_{b_j}^l} \prod_{i_1 \in \Phi^k_l} \mathcal{P}_f^{i_1j} \prod_{i_2 \in \mathcal{S}_j^{U} \backslash \Phi^k_l} \mathcal{\bar P}_f^{i_2j} \label{eq1_css_2}
\eeqn
where $\Phi^k_l$ in (\ref{eq1_css_1}) and (\ref{eq1_css_2}) are particular sets with $l$ SUs whose sensing outcomes indicate that 
channel $j$ is busy given that this channel is indeed busy and idle, respectively; ${\vec \varepsilon ^j} = \left\{\varepsilon ^{ij} \right\}$, ${\vec \tau^j} = \left\{\tau^{ij}\right\}$, $i \in \mathcal{S}_j^U$. 
For brevity, $\mathcal{P}_d^j\left({\vec \varepsilon ^j}, {\vec \tau^j}, a_j  \right) $ and  $\mathcal{P}_f^j\left( {\vec \varepsilon ^j}, {\vec \tau^j}, a_j  \right)$ are written as $\mathcal{P}_d^j$ and $\mathcal{P}_f^j$ in the following.

\subsection{MAC Protocol Design}
\label{MACDesign}

\begin{figure}[!t] 
\centering
\includegraphics[width=60mm]{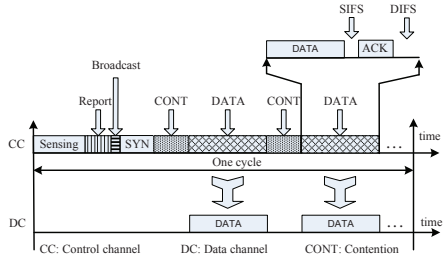}
\caption{Timing diagram of the proposed multi-channel MAC protocol }
\label{Fig2}
\end{figure}

We assume that time is divided into fixed-size cycles and it is assumed that
SUs can perfectly synchronize with each other (i.e., there is no synchronization error) \cite{Konda08}.
We propose a synchronized multi-channel MAC protocol for dynamic spectrum sharing as follows.
The MAC protocol has three phases in each cycle utilizing one control channel, which is assumed to be always available, as illustrated in Fig.~\ref{Fig2}.
In the first phase, namely the sensing phase of length $\tau$, all SUs simultaneously perform spectrum sensing on
their assigned channels. Here, we have $\tau = \max_{i} \tau_i$, where $\tau_i = \sum_{j \in \mathcal{S}_i} \tau_{ij}$ is total sensing
time of SU $i$, $\tau_{ij}$ is the sensing time of SU $i$ on channel $j$, and $\mathcal{S}_i$ is the set of channels assigned for SU $i$. 
All SUs exchange beacon signals on the control channel to achieve synchronization
in the second phase. Then, each SU reports its sensing results to the AP on the control channel. The AP collects sensing results from all SUs;
decide idle/status for all channels; and broadcast this information to all SUs on the control channel.

In the third phase, SUs participate in the contention and 
the winning SU will transmit data on all vacant channels. 
We assume that the length of each cycle is sufficiently large so that SUs can transmit several packets during the data transmission phase. 
During the data transmission phase, we assume that active SUs employ a standard contention technique to capture the channel 
similar to that in the CSMA/CA protocol.
Exponential backoff with minimum contention window $W$ and maximum backoff stage $m_0$ \cite{R3} is employed in the contention phase.
For brevity, we refer to $W$ simply as contention window in the following.
Specifically, suppose that the current backoff stage of a particular SU is $i$ then it starts the contention by choosing a random backoff time uniformly distributed in the range $[0,2^i W-1]$, $0 \leq i \leq m_0$. 
This user then starts decrementing its backoff time counter while carrier sensing transmissions from other SUs on vacant channels. 

Let $\sigma$ denote a mini-slot interval, each of which corresponds one unit of the backoff time counter. 
Upon hearing a transmission from any SU, each SU will ``freeze'' its backoff time counter and reactivate when the channel is sensed idle again. 
Otherwise, if the backoff time counter reaches zero, the underlying SU wins the contention. 
Here, two-way handshake will be employed to transmit one data packet on the available channel. 
After sending the data packet the transmitter expects an acknowledgment (ACK) from the receiver to indicate a successful reception of the packet. Standard small intervals, namely DIFS and SIFS, are used before backoff time decrements and ACK packet transmission as described in \cite{R3}.

\section{Performance Analysis, Design, and Optimization}
\label{MultipleChan}

\subsection{Throughput Analysis}
\label{ThroughputAO}

We assume that all SUs transmit data packets of the same length. 
Let $\mathcal{E}$ denote the average number of vacant channels that are correctly detected by the AP.
Suppose $\mathcal{T}\left( \tau, W  \right)$ denote the throughput achieved by all $N$ SUs on an imaginary single-channel network
where the channel is always available. 
Then, the normalized throughput per one channel achieved by our MAC protocol can be calculated as 
\beqn
\mathcal{NT} =  \mathcal{T}\left( \tau , W \right) \frac{1}{M} \mathcal{E} 
\eeqn
Here, $\mathcal{E}$ can be calculated as follows:
\beqn 
\mathcal{E} = \sum_{m =1}^M \sum_{i=1}^{C_M^{m }} \prod_{j_1 \in \Psi^{i}_m} \mathcal{P}_{j_1} \left(\mathcal{H}_0\right) \prod_{j_2 \in \mathcal{S} \backslash \Psi^{i}_m} \mathcal{P}_{j_2} \left(\mathcal{H}_1\right) \label{nput_1}\\
\times \sum_{n=1}^{m} \sum_{i_1=1}^{C_{m}^{n}} n \prod_{j_3 \in \Theta^{i_1}_n} \mathcal{\bar P}_f^{j_3} \prod_{j_4 \in \Psi^{i}_m \backslash \Theta^{i_1}_n} \mathcal{P}_f^{j_4} \label{nput_2} 
\eeqn
where $\mathcal{S}$ is the set of all $M$ channels. The quantity (\ref{nput_1}) represents the probability that there are $m$ available
 channels, which may or may not be correctly detected by SUs and the AP. Here, $\Psi^i_m$ denotes a particular set of $m$ available channels
  whose index is $i$. The second quantity (\ref{nput_2}) describes the product of $n$ and the probability that there are $n$ available channels 
according to the sensing decision of the AP (so the remaining available channels are overlooked due to sensing errors) where $\Theta^{i_1}_n$ denotes 
the $i_1$-th set with $n$ available channels. 

In the following, we describe how to calculate $\mathcal{T}\left( \tau , W \right)$
by using the technique developed by Bianchi in \cite{R3}. In particular,
we approximately assume a fixed transmission probability $\phi$ in a generic slot time. 
Bianchi shows that this transmission probability can be computed from
the following two equations \cite{R3}
\beqn \label{phi}
\phi  = \frac{{2\left( {1 - 2p} \right)}}{{\left( {1 - 2p} \right)\left( {W + 1} \right) + Wp\left( {1 - {{\left( {2p} \right)}^{m_0}}} \right)}},
\eeqn
\beqn \label{p}
p = 1-\left(1-\phi\right)^{n-1},
\eeqn
where $m_0$ is the maximum backoff stage, $p$ is the conditional collision probability (i.e., the probability that a collision is observed when a data packet is transmitted on the channel). 
For our system, there are $N$ SUs participating in contention in the third phase, the probability  that
at least one SU transmits its data packet can be written as
\beqn
\label{eq8a} 
\mathcal{P}_t = 1 - \left( 1 - \phi  \right)^N.
\eeqn
However, the probability that a transmission occurring on the channel is successful given there is at least one SU
 transmitting can be written as
\beqn
\label{eq9a}
\mathcal{P}_s = \frac{N \phi \left( 1 - \phi  \right)^{N - 1}}{\mathcal{P}_t}.
\eeqn
The average duration of a generic slot time can be calculated as 
\beqn
\label{eq10a}
\bar T_{sd} = \left( 1 - \mathcal{P}_t \right) T_e + \mathcal{P}_t \mathcal{P}_s T_s + \mathcal{P}_t \left( 1 - \mathcal{P}_s \right) T_c,
\eeqn
where $T_e = \sigma$, $T_s$ and $T_c$ represent the duration of an empty slot, the average time the channel is sensed busy due to a successful transmission, and the average time the channel is sensed busy due to a collision, respectively. These quantities can be calculated under 
the basic access mechanism as \cite{R3}
\beqn
\label{eq11a}
\left\{ \!\!\!{\begin{array}{*{20}{c}}
   T_s \!= \! T_s^{1}\! = \! H \!+ \! PS \!+ \! SIFS\! + \!2PD \!+ \!ACK \!+\! DIFS \hfill  \\
   T_c \!= \! T_c^{1}\! = \! H \!+ \! PS\! + \! DIFS \!+\! PD \hfill  \\
   H \!=\! H_{\text{PHY}} \!+ \! H_{\text{MAC}} \hfill  \\
\end{array}} \right.\!\!\!\!,
\eeqn
where $H_{\text{PHY}}$ and $H_{\text{MAC}}$ are the packet headers for physical and MAC layers, $PS$ is the packet size in transmission time, which is assumed to be fixed in this paper, $PD$ is the propagation delay, $SIFS$ is the length of a short interframe space, $DIFS$ is the length of a distributed interframe space, $ACK$ is the length of an acknowledgment. Recall that these parameters are measured in units of bits or $\mu s$ due to bit rate = 1 Mbps.
Based on these quantities, we have
\begin{equation}
\label{eq13a}
\mathcal{T} \left( \tau ,W  \right) = \left \lfloor \frac{T - \tau - T_R}{\bar T_{sd}} \right \rfloor \frac{\mathcal{P}_s \mathcal{P}_t PS}{T},
\end{equation}
where $T_R = Nt_r+t_b$, $t_r$ is the report time from each SU to the AP, $t_b$ the broadcast time from the AP to all SUs. 
Recall that $\tau = \max_{i} \tau^i$ is the total the sensing time.
$\left\lfloor  .  \right\rfloor $ denotes the floor function and recall that $T$ is the duration of a cycle. Note
that $\left\lfloor \frac{T - \tau - T_R}{\bar T_{sd} } \right\rfloor$ denotes the average number of generic slot times
in one particular cycle excluding the sensing and reporting phase. Here, we omit the length of the synchronization phase, which is assumed to be negligible.

\subsection{Cooperative Sensing and Access Optimization}
\label{OpCMP}

We discuss optimization of cooperative sensing and access parameters to maximize the normalized
 throughput under sensing constraints for PUs.
In particular, the throughput maximization problem can be stated as follows:
\beqn
 \mathop {\max} \limits_{\tau^{ij} ,W} \quad \mathcal{NT} \left( \tau ,W \right)  \hspace{3.4cm} \label{prob1}\\ 
 \mbox{s.t.}\,\,\,\, \mathcal{P}_d^j \left( {\vec \varepsilon ^j}, {\vec \tau^j}, a_j  \right) \geq \mathcal{\widehat{P}}_d ^j, j \in \left[1, M\right] \hspace{1cm} \label{prob2}\\
 \quad \quad 0 < \tau^{ij}  \le {T},  \quad 0 < W \leq W_{\sf max}, \hspace{1cm} \label{prob3}
\eeqn
where $\mathcal{P}_d^j$ is the detection probability for channel $j$ at the AP,
$W_{\sf max}$ is the maximum contention window and recall that $T$ is the cycle interval.

\begin{algorithm}[h]
\caption{\textsc{Sensing and Access Optimization}}
\label{mainalg}
\begin{algorithmic}[1]

\STATE Assume we have the sets of all SU $i$, $\left\{\mathcal{S}_i\right\}$. Initialize $\tau^{ij}$, $j \in \mathcal{S}_i$.

\STATE For each integer value of $W \in \left[1, W_{\sf max}\right]$, find ${\bar \tau}^{ij}$ as 

\FOR  {$i = 1$ \text{to} $N$}

\STATE Fix all $\tau^{i_1j}$, $i_1 \neq i$.

\STATE Find optimal ${\bar \tau}^{ij}$ as ${\bar \tau}^{ij} = \mathop {\argmax} \limits_{0 < \tau^{ij} \leq T} \mathcal{NT}\left(\tau^{ij}, W\right)$.

\ENDFOR

\STATE The final solution $\left({\bar W}, {\bar \tau}^{ij}\right)$ is determined as $\left({\bar W}, {\bar \tau}^{ij}\right) = \mathop {\argmax} \limits_{W, {\bar \tau}^{ij}} \mathcal{NT} \left({\bar \tau}^{ij}, W\right)$.

\end{algorithmic}
\end{algorithm}

We propose a low-complexity algorithm (Alg. 1) to find an efficient solution for the optimization problem (\ref{prob1}, \ref{prob2}, \ref{prob3}). 
In particular, for each potential value of  $W \in \left[1, W_{\sf max}\right]$, we search for the
best $\tau^{ij}$ to maximize the total throughput. This is done by a sequential search technique. Then,
the final solution is determined by the best combination of $\tau^{ij}, W$ for different values of $W$.
Numerical results reveal that Alg. 1 can always find the optimal solution of the underlying problem. 

\subsection{Channel Assignment for Throughput Maximization}
\label{GACAP}

So far we have assumed a fixed channel assignment based on which SUs perform sensing.
In this section, we attempt to determine an efficient channel assignment solution by solving the following problem
\beqn
\label{eq11a}
\max \limits_{ \left\{\mathcal{S}_i\right\} } \mathcal{NT} \left({\bar \tau}^{ij}, {\bar W}, 
 \left\{\mathcal{S}_i\right\} \right)
\eeqn

\begin{algorithm}[h]
\caption{\textsc{Channel Assignment Algorithm}}
\label{ChanAA}
\begin{algorithmic}[1]

\STATE Run Alg. \ref{mainalg} for temporary assignments $\mathcal{S}_i = \mathcal{S}$, $i \in \left[1, N\right]$ to get $\left({\bar W}, {\bar \tau}^{ij}\right)$. 
Employ Hungarian algorithm \cite{Kuhn55} to determine the first channel assignment for each SU so that each channel is assigned to exactly one SU
where the cost of assigning channel $j$ to SU $i$ is ${\bar \tau}^{ij}$. This results in initial channel assignment sets $\left\{\mathcal{S}_i\right\}$
for different SU $i$. 

\STATE $\text{continue} := 1$, $k:=1$.

\WHILE {\text{continue} = 1}

\STATE Calculate the normalized throughput with optimized parameter setting by using Alg. 1 as $\mathcal{NT}_{th} = \mathcal{NT} \left({\bar \tau}^{ij}, {\bar W}, \mathcal{S}_i\right)$.

\STATE Each SU $i$ calculates the increase of throughput if it is assigned one further potential channel $j$ as $\Delta T_{ij} = \mathcal{NT} \left({\bar \tau}^{ij}, {\bar W}, \mathcal{S}_i^1\right) - \mathcal{NT}_{th}$ where $\mathcal{S}_i^1 = \mathcal{S}_i \cup j$ and ${\bar \tau}^{ij}, {\bar W}$ are determined
by using Alg. 1 for assignment sets  $\mathcal{S}_i^1$ and  $\mathcal{S}_l, \: l \neq i$.

\STATE Find the ``best'' assignment $\left({\bar i}, {\bar j}\right)$ as $\left({\bar i}, {\bar j}\right) = \mathop {\argmax} \limits_{i, j \in \mathcal{S} \backslash \mathcal{S}_i} \Delta T_{ij} $.

\IF {$\Delta T_{\bar i \bar j} >\delta$}

\STATE Assign channel $\bar j$ to SU $\bar i$ ($\mathcal{S}_i = \mathcal{S}_i^1 $).

\STATE $k = k+1$.

\ELSE

\STATE $\text{continue} :=0$

\ENDIF

\ENDWHILE


\IF {$k>1$}

\STATE Return to step 2.

\ELSE 

\STATE STOP Alg.

\ENDIF

\end{algorithmic}
\end{algorithm}

\subsubsection{Brute-force Search Algorithm}
\label{ComAna}

Since the possible number of channel assignments is finite, we can employ the 
brute-force search to determine the optimal channel assignment solution and its protocol parameters.
 This can be done by determining the best configuration parameters under each channel assignment (i.e., using Alg. 1) then
 comparing the throughput achieved by different channel assignments to find the best one.
 
We now quantify the complexity of this optimal brute-force search algorithm. The number of possible assignments is equal to the following: How many 
ways are there to fill 1/0 to the elements of an $N$x$M$ matrix. It can verify that the 
 number of ways is $2^{MN}$. Therefore, the complexity of the optimal brute-force search algorithm is $\mathcal{O}\left(2^{MN}\right)$. 
 Moreover, for each case, we must run Alg. 1 to determine the sensing and access parameters. 

\subsubsection{Low-complexity Algorithm}
\label{ComAna}

We propose a low-complexity algorithm to find an efficient channel assignment solution, which is described in Alg. 2.
In step 1, we run Hungarian algorithm to perform the first channel assignment for each SU $i$. The complexity of this operation can be upper-bounded by $\mathcal{O}\left(M^2N\right)$ (see \cite{Kuhn55} for more details). In each assignment in the loop (i.e., Steps 2-13), each SU $i$
calculates the increases of throughput due to different potential channel assignments, and selects the one resulting in the maximum increase.
Hence, the complexity involved in these assignments is upper-bounded by $MN$ since 
there are at most $M$ channels to choose for each of $N$ SUs. Also, the number of assignments to perform is upper bounded by $MN$. Therefore, 
the complexity of this loop is upper-bounded by $M^2N^2$. Suppose we run these assignments for $r$ times before the algorithm terminates. Therefore,
the complexity of Alg. 2 can be upper-bounded by $\mathcal{O}\left(M^2N+rM^2N^2\right) = \mathcal{O}\left(rM^2N^2\right)$, which is much lower than
 that of the brute-force search algorithm.

\vspace{10pt}
\section{Numerical Results}
\label{Results}


\begin{table*} 
\centering
\caption{Throughput vs Probability $\mathcal{P}_j\left(\mathcal{H}_0\right)$ (MxN=4x4)}
\label{table1}
\begin{tabular}{|c|c|c|c|c|c|c|c|c|c|c|c|}
\cline{3-12} 
\multicolumn{2}{c|}{} & \multicolumn{10}{c|}{$\mathcal{P}_j\left(\mathcal{H}_0\right)$}\tabularnewline
\cline{3-12} 
\multicolumn{2}{c|}{} & 0.1 & 0.2 & 0.3 & 0.4 & 0.5 & 0.6 & 0.7 & 0.8 & 0.9 & 1\tabularnewline
\hline 
 & Greedy & 0.0838 & 0.1677 & 0.2515 & 0.3353 & 0.4191 & 0.5030 & 0.5868 & 0.6707 & 0.7545 & 0.8383\tabularnewline
\cline{2-12} 
$\mathcal{NT}$ & Optimal & 0.0846 & 0.1692 & 0.2544 & 0.3384 & 0.4239 & 0.5082 & 0.5935 & 0.6769 & 0.7623 & 0.8479\tabularnewline
\cline{2-12} 
 & Gap (\%) & 1.0090 & 0.9187 & 1.1293 & 0.9142 & 1.1294 & 1.0261 & 1.1266 & 0.9159 & 1.0261 & 1.1266\tabularnewline
\hline
\end{tabular}
\end{table*}

To obtain numerical results, we take key parameters for the MAC protocols from Table II in \cite{R3}. Other parameters are chosen as follows:
cycle time is $T = 100 ms$; mini-slot (i.e., generic empty slot time) is $\sigma = 20 {\mu} s$;
sampling frequency for spectrum sensing is $f_s = 6 MHz$;  bandwidth of PUs' QPSK signals is $6 MHz$; $t_r = 80 {\mu} s$ and $t_b = 80 {\mu} s$. 
The target detection probabilities for channel $j$ $\mathcal{\widehat{P}}^j_d$ in (\ref{prob2}) are chosen randomly in the intervals $[0.95, 0.99]$. 
In order to calculate $\mathcal{E}$ in (\ref{nput_1})-(\ref{nput_2}), we need to determine $\mathcal{P}_f^{j}$ for different $j$, which
can be done as follows. We set the equality for (\ref{prob2}), i.e., $\mathcal{P}_d^j\left( {\vec \varepsilon ^j}, {\vec \tau^j} , a_j  \right) = \mathcal{\widehat{P}}_d ^j$ (see \cite{Le11} for detailed explanation) and assume that detection probabilities $\mathcal{P}_d^{ij}=\mathcal{{P}}_d^{j*}$
are equal to each other from which we can calculate $\mathcal{{P}}_d^{j*}$ by using (\ref{eq1_css_1}). Then, we can determine $\mathcal{P}_f^{j}$ by using
(\ref{eq1_css_2}) and (\ref{eq2}). 
The signal-to-noise ratio of PU signals at SUs $\gamma^{ij}=SNR_p^{ij}$ are chosen randomly in the range $[-15, -20] dB$ and
the maximum backoff stage is  $m_0 = 3$.

We first compare the throughput performance achieved by the brute-force search and low-complexity algorithms (i.e., Alg. 2) for
channel assignment. In particular, in Table ~\ref{table1} we show the normalized throughput $\mathcal{NT}$ versus probabilities $\mathcal{P}_j\left(\mathcal{H}_0\right)$ for these two algorithms. Here, the probabilities $\mathcal{P}_j\left(\mathcal{H}_0\right)$ for
different channels $j$  are chosen to be the same and we choose $M = 4$ channels and $N = 4$ SUs. This figure confirms that the throughput gaps
between our greedy algorithm and the brute-force optimal search algorithm are quite small, which is about  1\% in all cases. 
These results confirm that our proposed greedy algorithm works well for small systems (i.e., small $M$ and $N$).


\begin{figure}[!t]
\centering
\includegraphics[width=60mm]{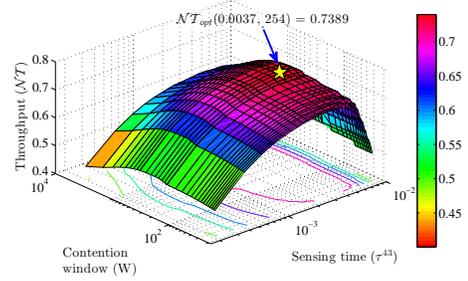}
\caption{Normalized throughput versus contention window $W$ and sensing time $\tau $ for $m = 3$, $N=10$, $M=4$.}
\label{Fig4}
\end{figure}

\begin{figure}[!t]
\centering
\includegraphics[width=60mm]{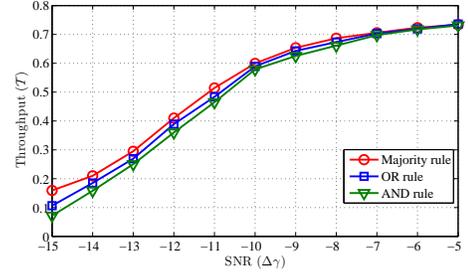}
\caption{Normalized throughput versus SNR shift $\Delta \gamma$ for $m = 3$, $N=10$, $M=4$ under 3 aggregation rules.}
\label{Fig5}
\end{figure}


\begin{figure}[!t]
\centering
\includegraphics[width=60mm]{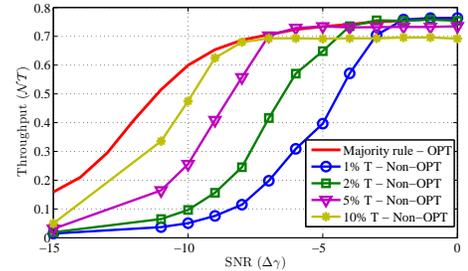}
\caption{Normalized throughput versus SNR shift $\Delta \gamma$ for $m = 3$, $N=10$, $M=4$ for optimized and non-optimized scenarios.}
\label{Fig6}
\end{figure}

\begin{figure}[!t]
\centering
\includegraphics[width=60mm]{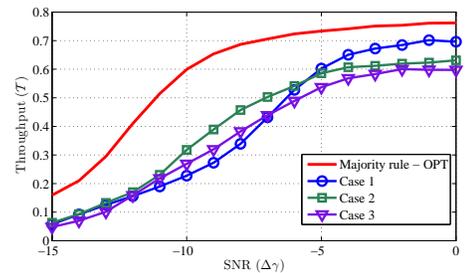}
\caption{Normalized throughput versus SNR shift $\Delta \gamma$ for $m = 3$, $N=10$, $M=4$ for optimized and RR channel assignments.}
\label{Fig7}
\end{figure}

 We now investigate the performance of our proposed
algorithm for larger systems. In Figs.~\ref{Fig4},~\ref{Fig5},~\ref{Fig6}, and \ref{Fig7}, we consider the network setting with $N=10$ and $M=4$.
We divide SUs into 2 groups where SUs have received SNRs due to PU $i$'s signal equal to $\gamma^{ij} = -15 dB$ and $\gamma^{ij} = -10 dB$ in the two
groups, respectively. 
We use a combination $\left\{ i,j \right\}$ to represent the scenario where channel $j$ is assigned to and sensed by SU $i$. The following combinations are set
corresponding to $\gamma^{ij} = -10 dB$: channel 1: $\left\{1,1\right\}, \left\{2,1\right\}, \left\{3,1\right\}$; channel 2: $\left\{2,2\right\}, \left\{4,2\right\}, \left\{5,2\right\}$; channel 3: $\left\{4,3\right\}, \left\{6,3\right\}, \left\{7,3\right\}$; and channel 4: $\left\{1,4\right\}, \left\{3,4\right\}, \left\{6,4\right\}, \left\{8,4\right\}, \left\{9,4\right\}, \left\{10,4\right\}$. The remaining combinations correspond to 
the SINR value $\gamma^{ij} = -15 dB$. To obtain results for different values of SNRs, we shift both SNRs (-10dB and -15dB) by  
$\Delta \gamma$. For example, when $\Delta \gamma = -10$, the resulting SNR values are $\gamma^{ij} = -25 dB$ and $\gamma^{ij} = -20 dB$. 

Fig.~\ref{Fig4} presents the normalized throughput $\mathcal{NT}$ versus contention window $W$ and sensing time $\tau^{ij}$ for the
combination $\left\{4,3\right\}$ and $\Delta \gamma = -5$ (the parameters of other combinations are set at optimal values). 
Therefore, this figure shows the normalized throughput $\mathcal{NT}$ versus $W$ and only $\tau^{43}$. 
We show the optimal configuration $\left({\bar \tau}^{43}, {\bar W}\right)$, which maximizes the normalized 
throughput $\mathcal{NT}$ of the proposed  MAC protocol. 
It can be observed that the normalized throughput $\mathcal{NT}$ is less sensitive to the contention window $W$ 
while it decreases significantly as the sensing time $\tau^{43}$ deviates from the optimal value.

In Fig.~\ref{Fig5}, we compare the  throughput performance as the AP employs there different aggregation rules, namely AND, OR, and Majority rules. The
three throughput curves in this figure represent the optimized normalized throughputs (i.e., by using Algs. 1 and 2). It can be seen that 
The Majority rule achieves the highest throughput among the three. In Fig.~\ref{Fig6}, we compare the throughput performances under the optimized
and the non-optimized scenarios. For the non-optimized scenario, we also employ Alg. 2 for channel assignment; however we do not use Alg.1 
to choose optimal sensing/access parameters. Instead, $\tau^{ij}$ is chosen from the following values: $1\% T$, $2\%T$, 
$5\%T$ and $10\%T$, where $T$ is the cycle time. Again, the optimized normalized throughput is higher than that due to non-optimized scenarios.
Finally, Fig. ~\ref{Fig7} demonstrates the relative throughput performance of our proposed algorithm and the round-robin (RR) channel assignment strategies. 
For RR channel assignments, we allocate channels for users, which is described in Table~\ref{table_round}. 
For all RR channel assignments, we employ Alg. 1 to determine optimal sensing/access parameters. 
Again, the optimized design achieve much higher throughput than those due to RR channel assignments.

\begin{table}
\centering
\caption{Round-robin Channel Assignment (x denotes an assignment)}
\label{table_round}
\begin{tabular}{|c|}
\multicolumn{1}{c}{}\tabularnewline
\multicolumn{1}{c}{}\tabularnewline
\hline 
1\tabularnewline
\hline 
2\tabularnewline
\hline 
3\tabularnewline
\hline 
4\tabularnewline
\hline 
5\tabularnewline
\hline 
6\tabularnewline
\hline 
7\tabularnewline
\hline 
8\tabularnewline
\hline 
9\tabularnewline
\hline 
10\tabularnewline
\hline 
\end{tabular}%
\begin{tabular}{|c|c|c|c|}
\hline 
\multicolumn{4}{|c}{Case 1}\tabularnewline
\hline 
1 & 2 & 3 & 4\tabularnewline
\hline 
x &  &  & \tabularnewline
\hline 
 & x &  & \tabularnewline
\hline 
 &  & x & \tabularnewline
\hline 
 &  &  & x\tabularnewline
\hline 
x &  &  & \tabularnewline
\hline 
 & x &  & \tabularnewline
\hline 
 &  & x & \tabularnewline
\hline 
 &  &  & x\tabularnewline
\hline 
x &  &  & \tabularnewline
\hline 
 & x &  & \tabularnewline
\hline 
\end{tabular}%
\begin{tabular}{|c|c|c|c|}
\hline 
\multicolumn{4}{|c}{Case 2}\tabularnewline
\hline 
1 & 2 & 3 & 4\tabularnewline
\hline 
x & x &  & \tabularnewline
\hline 
 & x & x & \tabularnewline
\hline 
 &  & x & x\tabularnewline
\hline 
 &  &  & x\tabularnewline
\hline 
x & x &  & \tabularnewline
\hline 
 & x & x & \tabularnewline
\hline 
 &  & x & x\tabularnewline
\hline 
 &  &  & x\tabularnewline
\hline 
x & x &  & \tabularnewline
\hline 
 & x & x & \tabularnewline
\hline 
\end{tabular}%
\begin{tabular}{|c|c|c|c|}
\hline 
\multicolumn{4}{|c}{Case 3}\tabularnewline
\hline 
1 & 2 & 3 & 4\tabularnewline
\hline 
x & x & x & \tabularnewline
\hline 
 & x & x & x\tabularnewline
\hline 
 &  & x & x\tabularnewline
\hline 
 &  &  & x\tabularnewline
\hline 
x & x & x & \tabularnewline
\hline 
 & x & x & x\tabularnewline
\hline 
 &  & x & x\tabularnewline
\hline 
 &  &  & x\tabularnewline
\hline 
x & x & x & \tabularnewline
\hline 
 & x & x & x\tabularnewline
\hline 
\end{tabular}
\end{table}

\vspace{0.2cm}
\section{Conclusion}
\label{conclusion} 

We propose a general analytical framework for cooperative sensing and access design and optimization in cognitive radio networks. We 
analyze the throughput performance of the proposed design, and develop an algorithm to find its sensing/access parameters. 
Moreover, we  present both optimal brute-force search and low-complexity algorithms to determine efficient channel assignments. Then,
 we analyze the complexity of different algorithms and evaluate their throughput performance via numerical studies.



\bibliographystyle{IEEEtran}


\end{document}